\documentclass[journal=jctcce,manuscript=article]{achemso}

\usepackage{chemformula} 
\usepackage[T1]{fontenc} 
\usepackage{bigstrut,multirow,rotating,booktabs}
\usepackage{threeparttable}
\usepackage{graphicx}
\usepackage{float}
\usepackage{amsmath}
\usepackage{amssymb}
\usepackage{subfigure}
\usepackage{braket}
\usepackage{afterpage}
\usepackage{diagbox}
\usepackage{multirow}
\usepackage{xr}
\usepackage{makecell}
\usepackage{caption}
\definecolor{ao(english)}{rgb}{0.0, 0.42, 0.24}

\author{Xinyuan Liang}
\affiliation{Academy for Advanced Interdisciplinary Studies, Peking University, Beijing, 90871, P. R. China}
\alsoaffiliation{HEDPS, CAPT, College of Engineering, Peking University, Beijing, 100871, P. R. China}
\author{Renxi Liu}
\affiliation{Academy for Advanced Interdisciplinary Studies, Peking University, Beijing, 90871, P. R. China}
\alsoaffiliation{HEDPS, CAPT, College of Engineering, Peking University, Beijing, 100871, P. R. China}
\author{Mohan Chen}
\alsoaffiliation{Academy for Advanced Interdisciplinary Studies, Peking University, Beijing, 90871, P. R. China}
\alsoaffiliation{HEDPS, CAPT, College of Engineering, Peking University, Beijing, 100871, P. R. China}
\email{mohanchen@pku.edu.cn}

\title{Investigating CO Adsorption on  Cu(111) and Rh(111) Surfaces Using Machine Learning Exchange-Correlation Functionals}

\begin{document}




\begin{abstract}

The ``CO adsorption puzzle'', a persistent failure of utilizing generalized gradient approximations (GGA) in density functional theory to replicate CO's experimental preference for top-site adsorption on transition-metal surfaces, remains a critical barrier in surface chemistry.
While hybrid functionals such as HSE06 partially resolve this discrepancy, their prohibitive computational cost limits broader applications. 
We tackle this issue by adopting the Deep Kohn-Sham (DeePKS) method to train machine-learned exchange-correlation functionals.
Principal component analysis reveals that the input descriptors for electronic structures separate distinctly across different chemical environments, enabling the DeePKS models to generalize to multi-element systems.
We train system-specific DeePKS models for transition-metal surfaces Cu(111) and Rh(111). 
These models successfully recover experimental site preferences, yielding adsorption energy differences of about 10 meV compared to HSE06.
Furthermore, a single model for the two surfaces is trained, and the model achieves comparable accuracy in predicting not only adsorption energies and site preference but also potential energy surfaces and relaxed surface adsorption structures.
%
The above work demonstrates a promising path towards universal models, enabling catalyst exploration with hybrid functional accuracy at substantially reduced cost.
%

\end{abstract}


$\\$
{\bf Keywords}
$\\$
CO adsorption; Transition Metal Surface; Machine-learning; Exchange-correlation functional; DeeP Kohn-Sham; Site preference; Transferability  

\section{Introduction}

Density functional theory (DFT)~\cite{64PR-Hohenberg,65PR-Kohn} serves as the cornerstone for investigating surface properties like adsorption mechanisms, yet the accuracy critically depends on the exchange-correlation functional.
The approximation forms of the exchange-correlation functional constitute the so-called Jacob’s ladder,~\cite{01AIP-Perdew} where higher rungs representing higher physical accuracy at the expense of increased computational cost.
Within the Jacob's ladder hierarchy, widely-used Perdew-Burke-Ernzerhof (PBE)~\cite{96PRL-Perdew} functional, a generalized gradient approximation at the second rung, achieves a favorable balance between computational efficiency and accuracy for extended systems, albeit with inherent and persistent delocalization errors.
Occupying a higher rung on Jacob's ladder, hybrid functionals, such as HSE06~\cite{03JCP-Heyd,06JCP-Krukau} and PBE0~\cite{96JCP-Perdew}, incorporate a fraction of exact exchange that partially resolves the electron delocalization errors.
However, this accuracy comes at a severe computational cost, as the evaluation of exact exchange integrals scales $O(N^4)$ with system size, rendering hybrid functionals prohibitively expensive for routine application to large-scale systems.

A typical example of such differences arising from exchange-correlation functionals can be seen in interfacial phenomena, particularly the ``CO adsorption puzzle''~\cite{01JPCB-Feibelman} on transition metal surfaces, most notably Pt(111), Rh(111), and Cu(111).
At low coverages, experimental observations consistently show CO's preference for top-site adsorption on these surfaces,~\cite{77SS-Kessler,82SS-Steininger,86SS-Ogletree,88SS-Raval} while GGA calculations incorrectly stabilize the higher-coordination face-centered cubic (fcc) site.~\cite{01JPCB-Feibelman,02JCP-Grinberg,03SurSci-Gil,19PRB-Patra}
The Blyholder model~\cite{64JPC-Blyholder,98SurSci-Aizawa} interprets CO chemisorption through interactions of CO frontier orbitals with metallic $d$-band.
However, PBE's inherent tendency to underestimate electronic band gaps leads to an artificial overstabilization of fcc adsorption sites due to spatial considerations.~\cite{95PRB-Illas,00JCP-Koper,03SurSci-Gil,04SurSci-Doll,18PCCP-Gameel}
Several strategies have been explored~\cite{02JCP-Grinberg,10PRB-Lazic} to address this issue.
Approaches including DFT+U method with orbital-specific empirical Coulomb corrections~\cite{03PRB-Kresse,04PRB-Kohler,05SurSci-Gajdos}, the RPBE functional featuring modified gradient approximations~\cite{99PRB-Hammer,04JPCM-Gajdo}, and the BEEF-vdW\cite{12PRB-Wellendorff,16PRB-Lundgaard} framework integrating non-local van der Waals interactions and empirical parametrization,~\cite{15SS-Wellendorff,22NC-Araujo} achieve tailored accuracy for adsorption site preferences but require empirical parameterization or compromise transferability.
Physically motivated advancements on Jacob's ladder provide alternative pathways.
Hybrid functionals such as HSE06 mitigate GGA limitations through its nonlocal exchange formalism.
This physically-motivated approach significantly improves accuracy, achieving experimental agreement for Cu(111) and Rh(111) systems.~\cite{07PRB-Stroppa,08NJP-Stroppa}
The random phase approximation (RPA)~\cite{12JMS-Ren} further improves absolute adsorption energy accuracy~\cite{09PRB-Ren,10NM-Schimka} by incorporating dynamic electron correlations, but at the cost of more prohibitive computational scaling.

In recent years, machine learning (ML) has emerged as an innovative approach to reconcile the long-standing trade-off between computational accuracy and efficiency.~\cite{23-Batatia}
For instance, ML force fields (MLFFs) achieving RPA accuracy enable systematic exploration of coverage-dependent CO adsorption behavior on the Rh(111) surface.~\cite{23PRL-Liu}
Furthermore, MLFFs facilitate rapid characterization of adsorption phenomena across diverse crystal facets~\cite{25JMI-Wu} and femtosecond laser-induced desorption process.~\cite{24JPCL-Muzas,24JACSAu-Zugec}
Leveraging ML-enhanced property prediction frameworks, researchers can efficiently screen molecular adsorption configurations,~\cite{24CS-Guo} extract physical insights through high-throughput calculations, and elucidate correlations between adsorption energetics and critical electronic parameters.~\cite{22C-Agarwal}

Nevertheless, existing ML studies lack comprehensive investigation of CO adsorption mechanisms across varied transition-metal surfaces. 
To address this gap, we adopt the Deep Kohn-Sham (DeePKS)~\cite{21JCTC-Chen,22JPCA-Li} framework and propose a ML model capable of establishing accurate relative energy ordering for CO adsorption configurations across two 
 transition metal surfaces.
This ML-based exchange-correlation functional utilizes neural networks to resolve energy discrepancies between low- and high-level functionals, effectively capturing differences in total energies, atomic forces and etc. 
The resulting DeePKS formalism demonstrates seamless integration with self-consistent field (SCF) calculations, preserving the computational efficiency of low-level functionals like PBE while attaining accuracy of high-level functional like HSE06. 
Furthermore, DeePKS can effectively capture variations in electronic structure properties across multiple material systems. 
A universal DeePKS model for halide perovskites attains HSE06-level accuracy in band gap calculations,~\cite{23JPCC-Ou} while a DeePKS-ES (electronic structrue) model for molecular and liquid water systems successfully reproduces Hamiltonian matrices, band structures, and density of states with HSE06 accuracy.~\cite{25JCTC-Liang}

This work employed the DeePKS method as a machine-learned exchange-correlation functional in DFT to study CO adsorption on transition metals, including Cu(111) and Rh(111) surfaces. 
Dedicated models DeePKS-Cu/DeePKS-Rh were trained and validated for Cu(111)/Rh(111) adsorption systems, achieving optimal balance between chemical accuracy and computational efficiency. 
Significantly, we advance a model named DeePKS-Cu+Rh that simultaneously captures adsorption behaviors across both substrates, establishing a transferable framework for simulations of complex catalytic interfaces.

The rest of the paper is structured as follows. Section \ref{sec:methods} describes our computational methods. Section \ref{sec:results} presents the key results, including details for isolated CO molecule, the bare metal surfaces, CO adsorption enegies on Cu(111) and Rh(111) surfaces, and the corresponding potential energy surfaces. Finally, Section \ref{sec:conclusions} summarizes the main findings and conclusions.

\section{Methods}\label{sec:methods}

\subsection{Atomic Structures}
\begin{figure}[htbp]
  \includegraphics[width=16cm]{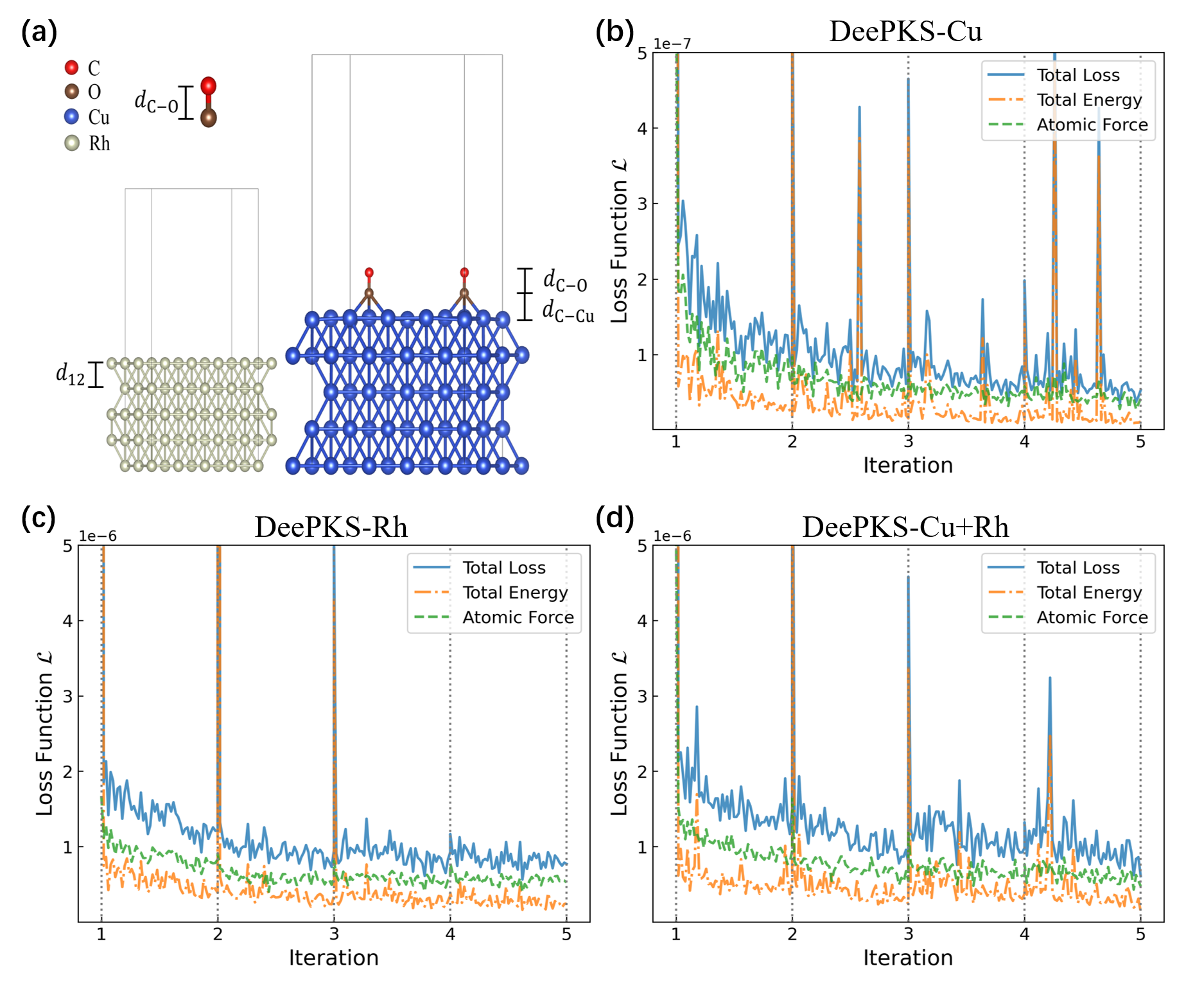}
  \caption{(Color online)
  (a) Examples of the atomic structures and length quantities involved in this work, including a gas-phase CO molecule, a clean Rh(111) surface, and a Cu(111) surface with CO adsorbed at the fcc site.
  (b-d) The training loss curves on the logarithmic scale for (b) DeePKS-Cu, (c) DeePKS-Rh, and (d) DeePKS-Cu+Rh model. The number on the horizontal axis indicates the initial training step for each iteration, while the vertical axis shows the values of the total loss function $\mathcal{L}$ (top blue curve) along with its components, including contributions from total energy and atomic forces.
  }\label{system_and_loss}
\end{figure}

We investigated three categories of atomic structures, which are isolated CO molecules, bare surfaces of Cu(111) and Rh(111), and CO-adsorbed surfaces.
Fig.~\ref{system_and_loss}(a) illustrates representative examples with their associated length-scale parameters.
For the gas-phase CO molecule, the equilibrium bond length is defined as $d_\text{C-O}$.
The clean Cu(111) and Rh(111) surfaces are modeled as five-layer slabs with a 2×4 supercell containing 40 metal atoms per unit cell. 
A vacuum region of 15 \AA~is applied along the $z$-direction to separate periodic images.
An example of the bare Rh(111) surface is displayed in Fig.~\ref{system_and_loss}(a), where the interlayer distance between the top two layers is denoted as $d_{12}$. 
For CO adsorption, two CO molecules are adsorbed on one side of the slab, resulting in a c(2×4) supercell with 44 atoms in total.
Fig.~\ref{system_and_loss}(a) illustrates the Cu(111) surface with CO molecule adsorbed at the fcc site, where the C-O and C-Cu bond lengths are labeled as $d_{\text{C-O}}$ and $d_{\text{C-Cu}}$, respectively.

\subsection{Descriptor of Electronic Structure}

The DeePKS method employs a neural network to model the energy difference $E^{\delta}$, which is calculated as the difference between the target energy $E^{\mathrm t}$ and the base energy $E^{\mathrm b}$. In this regard, the total energy given by a DeePKS model is 
\begin{equation}
E^{\mathrm d}=E^{\mathrm b}+E^{\delta}.
\end{equation}
Here the second term $E^{\delta}$ is decomposed into atomic contributions as
\begin{align}
E^\delta=\sum_{I}F_{\mathrm{NN}}(\mathbf{d}^I|\omega),
\end{align}
where $F_{\mathrm{NN}}$ denotes the neural network with parameters $\omega$. The input descriptor $\mathbf{d}^I$ encodes atom-centered electronic structure features for atom $I$.
Specifically, the electronic structure descriptor is constructed by first projecting the density matrix $\rho_{\mu\nu}$ of the basis $|\phi_{\mu}\rangle$ onto localized orbitals $|\alpha_{nlm}^{I}\rangle$,
\begin{equation}\begin{aligned}
D_{{nlmm}^{\prime}}^I&=
\sum_{\mu\nu}\rho_{\mu\nu}\langle\phi_{\mu}|\alpha_{{nlm}}^I\rangle\langle\alpha_{{nlm}^{\prime}}^I|\phi_{\nu}\rangle.
\end{aligned}\end{equation}
$n$, $l$, $m$ represent principal, angular momentum and magnetic quantum number, respectively.
Sub-blocks of the matrix $D_{{nlmm}^{\prime}}^I$ sharing identical $I$, $n$, $l$ indices are then diagonalized. 
After diagonalization, all eigenvalues associated with atom $I$ are then collected to form the atomic descriptor $\mathbf{d}^I$, thus capturing its chemical environment through a symmetry-adapted representation of the local electron density.
Moreover, the forces acting on atom $I$ can be computed by the given total energy from the DeePKS method
\begin{equation}
\mathbf{F}^\mathrm{d}_{I}=-\frac{\partial E^\mathrm{d}}{\partial \tau_{I}},
\end{equation}
where $\tau_I$ denotes the coordinate of atom $I$.

\subsection{Loss Function}

We trained DeePKS models by minimizing the loss function of total energies and atomic forces. 
Subject to the electronic wavefunction orthogonality constraint $\langle\psi_i|\psi_j\rangle=\delta_{ij}$, the training purpose is expressed as
\begin{equation}
\begin{aligned}
\label{eq:two-min}
\min_{\boldsymbol{\omega}}\mathbb{E}_{\text{data}}\Big[\Big({E^{\mathrm t}-E^{\mathrm d}[\{\widetilde{\psi}_i\}|\omega]}\Big)^2 +\frac{\sum_{I=1}^{N_{\rm a}}\Vert\mathbf{F}_{I}^{\mathrm t}-\mathbf{F}_{I}^{\mathrm d}[\{\widetilde{\psi}_i\}|\omega]\Vert^2}{3N_{\rm a}} \Big] ,
\end{aligned}
\end{equation}
where the superscripts $\mathrm t$ and $\mathrm d$ denote the target functional and the DeePKS model results, respectively.
Here $\mathbb{E}_{\text{data}}$ averages over training data, and $N_{a}$ is the number of atoms.
Note that $E^{\mathrm d}$ and $F^{\mathrm d}$ are computed through SCF calculations, where the optimized wavefunctions $\{\widetilde{\psi}_i\}$ satisfy $\{\widetilde{\psi}_i\}=argmin_{\{{\psi}_i\}}E^{\mathrm d}$.
Thus, $E^{\mathrm d}$ and $F^{\mathrm d}$ depend on both wave functions ${\{\psi_i\}}$, which define the descriptors $\mathbf{d}^I$, as well as model parameters $\omega$.

To reduce the number of SCF calculations while optimizing the parameters $\omega$, the DeePKS method employs an iterative training approach with alternating fixed-variable steps until convergence.~\cite{21JCTC-Chen,25JCTC-Liang}
Each iteration consists of two steps. 
In the first step, the neural network parameters $\omega$ remain fixed while solving the Kohn-Sham equations through SCF calculations. 
This generates ground-state wavefunctions and corresponding atomic descriptors $\mathbf{d}^I$.
In the second step, $\mathbf{d}^I$ remain frozen while updating the network parameters $\omega$ over multiple training epochs.
This workflow avoids recalculating wavefunctions during each parameter adjustment, significantly lowering SCF computations.

To illustrate the iterative training process and provide initial demonstration of model performances, Figs.~\ref{system_and_loss}(b-d) show changes of the loss function (Eq.~\ref{eq:two-min}) during four iterations when training the (b) DeePKS-Cu, (c) DeePKS-Rh, and (d) DeePKS-Cu+Rh models.
In all cases, both the loss terms of the total energy and atomic forces, as well as the total loss function, generally decrease within the training steps and between the iterations. 
While the total energy term occasionally exhibits large values, it typically remains small and eventually converges to stable solutions.
For the DeePKS-Cu model, the vertical axis scale in Fig.~\ref{system_and_loss}(b) is 1.0$\times$10$^{-7}$, which is smaller than 1.0$\times$10$^{-6}$ of the other two models in Figs.~\ref{system_and_loss}(c) and (d), 
resulting in substantially smaller loss values displayed.
The final converged losses for total energy and atomic forces reach magnitudes of $1\times10^{-8}$ and $5\times10^{-8}$, respectively. 
These losses of total energy and atomic forces result in different values of approximately 3 meV and 0.01 eV/\AA, respectively.
Impressively, the DeePKS-Rh and DeePKS-Cu+Rh models achieved comparable final accuracy, with total energy and atomic force errors reaching 10 meV and 0.03 eV/\AA, respectively.
Despite the increased complexity from more training data, the DeePKS-Cu+Rh model exhibits minimal precision loss.

\subsection{Principal Component Analysis for Descriptors}

\begin{figure}
  \includegraphics[width=10cm]{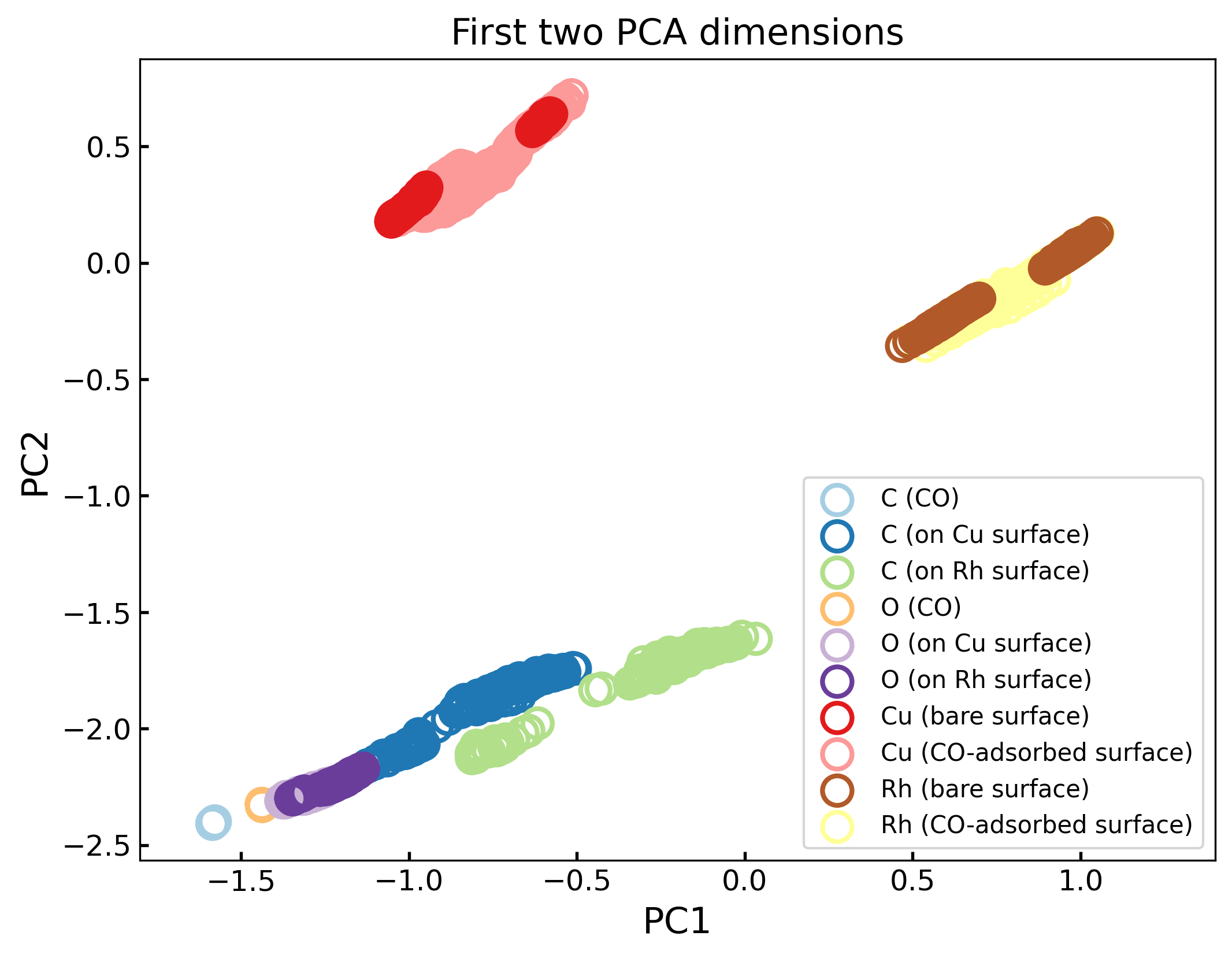}
  \caption{(Color online)
  Visualization of two-dimensional principal component analysis (PCA) for descriptors of all involved atoms. 
  The horizontal and vertical coordinates represent the first and second principal components (PC1 and PC2), respectively. 
  Different elements are distinguished in the plot, with parentheses indicating the different chemical environments of the corresponding atoms. 
  The data points include C and O atoms from both gas-phase CO molecules and CO adsorbed on Cu(111) or Rh(111) surfaces, along with Cu and Rh from bare metal surfaces and CO-adsorbed surfaces.
  }\label{pca}
\end{figure}

To investigate the discriminative capability of descriptors, we performed principal component analysis (PCA) on the atomic descriptors $\mathbf{d}^I$. 
The first two principal components (PC) collectively account for $93.5\%$ of the total variance, effectively capturing the essential different patterns among descriptors. 
Fig.~\ref{pca} plots the projected descriptor distributions along the first two PC axes (PC1/PC2), with data points color-coded according to elemental species and chemical environments. 
The legend explicitly annotates each element with its chemical environment in parentheses.
For instance, carbon atoms in CO molecules can be found either as isolated gas-phase molecules ``C (CO)'' or adsorbed on Cu ``C (on Cu surface)'' or Rh metal surfaces ``C (on Rh surface)''.

We find the PCA analysis of descriptors reveals three critical patterns in descriptor space segregation.
First, CO-derived descriptors exhibit distinct spatial separation from both Cu and Rh metal descriptors across the PC1/PC2 coordinates, reflecting the inherent elemental specificity in the descriptor design scheme.
%
Second, bare metal surfaces and their CO-adsorbed counterparts show overlapping descriptor distributions, likely because CO adsorption primarily modifies the electronic states of directly coordinated metal atoms while leaving others relatively unchanged.
Third, the descriptor distributions exhibit distinct environmental dependence for CO molecules.
The descriptor clusters show large spatial separation between gas-phase CO, Cu-adsorbed CO, and Rh-adsorbed CO systems. 
Meanwhile, gas-phase CO demonstrates limited descriptor diversity, forming a compact cluster in the PC space. 
In stark contrast, adsorbed CO configurations show significantly broadened descriptor distributions, with the spread of nearly 1 PC unit.
This is related to the fact that different adsorption sites (e.g., top vs. fcc) introduce additional variations in descriptor positions.
Particularly, the situation of separation and broadening is pronounced for the directly coordinating carbon atoms, whose descriptor clusters show complete spatial separation between different adsorption states.

Rooted in the electronic density distribution surrounding atoms, the descriptors effectively discriminate between distinct chemical environments of bare metal surfaces, isolated CO molecules, and CO adsorbed on different metallic surfaces, as demonstrated by their non-overlapping spatial distributions in the PCA analysis.
While this discriminative power enables precise identification of substrate-specific adsorption states, the observed lack of descriptor space overlap between different metallic systems and their adsorbed carbon atoms creates inherent transferability barriers. In addition, we find the DeePKS models trained exclusively on one metal system (e.g., DeePKS-Cu) cannot be directly transferred to other metal systems (e.g., Rh), also demonstrated by the failed generalization shown in Table~\ref{tab:Rh_adsorb}.
Consequently, hybrid training strategies incorporating more metal systems become essential for achieving cross-element transferability while preserving discriminative power, as demonstrated by the superior performance of the DeePKS-Cu+Rh model.

\subsection{Computational Details}

All of the DFT calculations were performed by the ABACUS\cite{10JPCM-Chen,16CMS-Li} package v3.8.0 with periodic boundary conditions.
A kinetic energy cutoff of 100 Ry was applied, together with the norm-conserving Vanderbilt (ONCV) pseudopotentials~\cite{15CPC-Schlipf} generated with the PBE exchange-correlation functional.
The calculations employed numerical atomic orbital (NAO) basis sets of double-zeta plus polarization (DZP) quality.
The raidus cutoffs were 8.0 Bohr for Cu and 7.0 Bohr for Rh, C and O.
Within the basis set and pseudopotential framework, the third and fourth electron shells (e.g., 3s, 3p, 3d, and 4s orbitals for Cu) were treated as valence electrons for both transition metals.

In PBE calculations, we adopted a charge density convergence criterion of $1.0\times10^{-6}$.
Charge mixing was performed using the Broyden method~\cite{88PRB-Johnson} with a mixing ratio of 0.2 for new charge density contributions.
HSE06 calculations were conducted using the LibRI v0.2.1 package, which implements the resolution of the identity (RI) method to significantly reduce computational costs with NAO basis sets. 
We specifically configured LibRI to use double-precision data types instead of complex numbers, thereby improving the efficiency of SCF calculations with hybrid functionals.
For HSE06 SCF iterations, convergence criteria required either a charge density difference below $1.0\times10^{-6}$ or an energy difference smaller than $5.0\times10^{-8}$ eV between successive iterations. 
Brillioun-zone sampling was done with a 8×6×1 $\Gamma$-centered $k$-point mesh.
During structural relaxation, the two uppermost surface layers and the CO molecule were allowed to move and the energy threshold was relaxed to $1.0\times10^{-7}$ eV.
The relaxation was achieved when all of the atomic forces were less then 0.04 eV/\AA.

When utilizing the DeePKS method, we employed spherical Bessel functions~\cite{10JPCM-Chen} as projected orbitals, with energy and radial cutoffs set to 100 Ry and 5.0 Bohr, respectively.
For the studied systems, we constructed fifteen sets of $s$-, $p$-, and $d$-orbitals, yielding a total of 135 projected localized orbitals.
We selected the PBE exchange-correlation functional as the base model and HSE06 as the target model, with total energy and atomic forces as the target properties.
HSE06 serves here as an example of computationally demanding and high-accuracy functionals.
Other functionals could equivalently be studied using this framework.
The neural network architecture in the DeePKS models consisted of a fully-connected multilayer perceptron containing 3 hidden layers of 100 neurons each.

A total of 90 and 100 structures were generated for the Cu and Rh systems, respectively, each including isolated CO molecules, bare metallic surfaces, and adsorbed surfaces when CO on four different sites (top, fcc, hcp, and bridge).
The number of structures in each category are detailed in Supplementary Table S1.
The relatively small size of the dataset effectively demonstrates the learning efficiency of the DeepKS method, highlighting its ability to achieve accurate predictions without requiring extensive training data.
For both Cu(111) and Rh(111) systems, we trained a specific DeePKS model, denoted as DeePKS-Cu and DeePKS-Rh, respectively. Additionally, we combined the above datasets to develop a model, named DeePKS-Cu+Rh, which works for both metal systems.
All models underwent four iterative training steps.

\section{Results and Discussion}\label{sec:results}

\subsection{CO Molecule and bare surface}

\begin{table}[htbp]
  \centering
  \caption{
  Bond length of CO molecule (denoted as $d_\text{CO}$), as well as the interlayer distances (in \AA) between the top two layers of the bare Cu(111) and Rh(111) surface (denoted as $d_{12}$).
  The values are calculated using different functionals, including the base functional PBE, the target functional HSE06, and three DeePKS models, along with reference values.
  }
  \label{tab:co_slab}
  \begin{threeparttable}
    \begin{tabular}{lcccccccc}
    \hline
    \hline
    \multirow{2}[4]{*}{Method} & \multicolumn{2}{c}{Bond lenth of CO} &       & \multicolumn{2}{c}{$d_{12}$ of Cu(111)} &       & \multicolumn{2}{c}{$d_{12}$ of Rh(111)} \bigstrut\\
\cline{2-3}\cline{5-6}\cline{8-9}          & $d_\text{C-O}$ & Ref.\tnote{a} &       & $d_{12}$ & Ref.\tnote{b} &       & $d_{12}$ & Ref.\tnote{b} \bigstrut\\
    \hline
    PBE   & 1.138  & 1.135  &       & 2.080  & 2.080  &       & 2.178  & 2.170  \bigstrut[t]\\
    HSE06   & 1.125  & 1.123  &       & 2.080  & 2.080  &       & 2.120  & 2.160  \\
    DeePKS-Cu & 1.125  & -     &       & 2.067  & -     &       & -     & - \\
    DeePKS-Rh & 1.125  & -     &       & -     & -     &       & 2.130  & - \\
    DeePKS-Cu+Rh & 1.125  & -     &       & 2.067  & -     &       & 2.122  & - \bigstrut[b]\\
    \hline
    \hline
    \end{tabular}%

\begin{tablenotes}
\footnotesize
\item[a] Ref.~\citenum{09PRB-Ren} (HSE06 reference values correspond to PBE0)
\item[b] Ref.~\citenum{07PRB-Stroppa} (HSE06 reference values correspond to HSE03)
\end{tablenotes}
\end{threeparttable}

\end{table}%

The structural analysis is performed for gas-phase CO and bare surface in Table~\ref{tab:co_slab}.
For the CO molecule, the experimental equilibrium C-O bond length~\cite{79-Huber} $d_\text{C-O}$ of 1.128 \AA~ is better reproduced by HSE06 (1.125 \AA), while PBE yields a slightly longer value (1.138 \AA).
These results agree with prior work of Ref.~\citenum{09PRB-Ren}, where PBE0 and PBE predicted $d_\text{C-O}$ as 1.123 and 1.135 \AA, respectively.
The DeePKS models demonstrate exceptional transferability, predicting relaxed C-O bond lengths within 0.001 \AA~error from HSE06 despite being trained on only 10 gas-phase CO configurations, highlighting their remarkable efficiency in capturing molecular structural features.
Regarding surface structural parameters, both PBE and HSE06 show reasonable agreement with reference $d_{12}$ values, which is the interlayer distance between the first and second atomic layers, except for slightly larger deviations of 0.04 \AA~for Rh surfaces by HSE06 calculations.
DeePKS predictions for $d_{12}$ in Cu systems show deviations of 0.013 \AA, likely due to insufficient data of only 10 bare surface configurations.
With more comprehensive training data of 20 bare surface configurations, DeePKS predictions in Rh systems demonstrate improved agreement of 0.01 \AA, highlighting the critical role of sufficient data sampling.

\subsection{Adsorption Energies of CO on Cu(111)}

\begin{table}[htbp]
  \centering
  \caption{
  Adsorption energies $E_\text{ads}$ (eV) for CO at the top site and fcc site on the Cu(111) surface calculated using different methods, including the base functional PBE, the target functional HSE06, and several DeePKS models.
  Adsorption energies are evaluated under two geometric conditions. First, relaxed geometry from the PBE functional. Second, relaxed geometry from the specific method used.
  Corresponding adsorption energy differences $\Delta E_{\rm ads}$ (top-fcc) and reference values are provided for systematic comparison.}
  \label{tab:Cu_adsorb}
    \begin{tabular}{clccc}
    \hline
    \hline
    \multicolumn{1}{c}{Geometry} & Method & $E_{\rm ads}(\rm top)$   & $E_{\rm ads}(\rm fcc)$   & $\Delta E_{\rm ads}$ \bigstrut\\
    \hline
    \multicolumn{1}{c}{\multirow{6}[2]{*}{\makecell{Relaxed geometry\\ from PBE}}} & PBE   & -0.741  & -0.850  & 0.109  \bigstrut[t]\\
          & PBE (Ref.~\citenum{07PRB-Stroppa}) & -0.72 & -0.86 & 0.14 \\
          & HSE06   & -0.573  & -0.488  & -0.086  \\
          & DeePKS-Cu & -0.572  & -0.489  & -0.084  \\
          & DeePKS-Cu+Rh & -0.580  & -0.491  & -0.089  \\
          & PBE0 (Ref.~\citenum{09PRB-Ren}) & -0.58  & -0.54  & -0.04  \bigstrut[b]\\
    \hline
    \multicolumn{1}{c}{\multirow{4}[2]{*}{\makecell{Relaxed geometry\\ from used method}}} & HSE06   & -0.581  & -0.506  & -0.075  \bigstrut[t]\\
          & DeePKS-Cu & -0.579  & -0.506  & -0.072  \\
          & DeePKS-Cu+Rh & -0.584  & -0.508  & -0.076  \\
          & HSE03 (Ref.~\citenum{07PRB-Stroppa}) & -0.561  & -0.555  & -0.006  \bigstrut[b]\\
    \hline
    \hline
    \end{tabular}%
\end{table}%

Table~\ref{tab:Cu_adsorb} systematically compares the differences in adsorption energy between CO molecules adsorbed at the top and fcc sites on the Cu surface obtained through different electronic structure methods. The geometries were obtained via a geometry relaxation process via two different methods, i.e., either from the PBE functional or from the employed electronic structure method.
First, by using the relaxed geometry from the PBE functional, electronic structure calculations with PBE reveals that the top site configuration exhibits a lower adsorption energy (i.e., less energy released upon adsorption) than the fcc site configuration, leading to a positive difference in adsorption energy ($\Delta E_{\rm ads}=E_{\rm ads}(\rm top)-E_{\rm ads}(\rm fcc)$) of 0.109 eV, indicating that the fcc site is the thermodynamically more stable configuration.
Both calculated adsorption energies and their differences show excellent agreement with Ref.~\citenum{07PRB-Stroppa}, with discrepancies maintained within 0.05 eV.
Unlike PBE, the hybrid functional HSE06 yields a negative energy difference of -0.086 eV based on PBE-relaxed geometries, aligning with experimental observations of top site preference.
The calculated adsorption energies exhibit deviations within 0.05 eV compared to Ref.~\citenum{09PRB-Ren} values obtained using the PBE0 functional.
When employing HSE06 self-consistently relaxed configurations, the energy difference decreases to -0.075 eV, but retains qualitative agreement with experiments.
The energy difference is considerably lower than that of Ref.~\citenum{07PRB-Stroppa}, which likely originates from a different basis selection.
The DeePKS-Cu model achieves remarkable consistency with HSE06, exhibiting only a difference of 2 meV in $\Delta E$ under PBE-optimized geometries, which are not included in the training dataset.
It maintains deviations lower than 1 meV/atom from HSE06 in several adsorption energy components, including gas-phase CO, bare surface, and adsorbed surface, as shown in Supplementary Table S2.
The DeePKS-Cu+Rh model, which is trained with Cu and Rh data simultaneously, shows slightly increased deviations of 3 meV while correctly reproducing energy rankings. 
In particular, both DeePKS models exhibit deviations of 2-3 meV from HSE06 in the adsorption energies even after self-consistent relaxations, maintaining qualitative agreement with the results of the experiments.
All of these suggest that the DeePKS models successfully encode the physical essence of HSE06 through machine learning, demonstrating robust generalization capabilities beyond the training dataset.

\subsection{Adsorption Energies of CO on Rh(111)}

\begin{table}[htbp]
  \centering
  \caption{
  Adsorption energies $E_\text{ads}$ (eV) for CO at the top site and fcc site on the Rh(111) surface calculated using different methods, including the base functional PBE, the target functional HSE06, and several DeePKS models.
  For HSE06 and DeePKS models, adsorption energies are evaluated under two geometric conditions. First, relaxed geometry from the PBE functional. Second, relaxed geometry from the specific method used.
  Corresponding adsorption energy differences $\Delta E_{\rm ads}$ (top-fcc) and reference values are provided for systematic comparison.}
  \label{tab:Rh_adsorb}
    \begin{tabular}{clccc}
    \hline
    \hline
    \multicolumn{1}{c}{Geometry} & Method & $E_{\rm ads}(\rm top)$   & $E_{\rm ads}(\rm fcc)$   & $\Delta E_{\rm ads}$ \bigstrut\\
    \hline
    \multicolumn{1}{c}{\multirow{6}[2]{*}{\makecell{Relaxed geometry\\ from PBE}}} & PBE   & -1.978  & -1.983  & 0.005  \bigstrut[t]\\
          & PBE (Ref.~\citenum{07PRB-Stroppa}) & -1.87 & -1.91 & 0.036 \\
          & HSE06   & -2.190  & -2.016  & -0.175  \\
          & DeePKS-Rh & -2.194  & -2.027  & -0.167  \\
          & DeePKS-Cu+Rh & -2.196  & -2.029  & -0.167  \\
          & DeePKS-Cu & -1.849  & -1.622  & -0.227  \bigstrut[b]\\
    \hline
    \multicolumn{1}{c}{\multirow{4}[2]{*}{\makecell{Method-relaxed\\ geometry}}} & HSE06   & -2.165  & -1.942  & -0.223  \bigstrut[t]\\
          & DeePKS-Cu & -2.177  & -1.947  & -0.230  \\
          & DeePKS-Cu+Rh & -2.183  & -1.943  & -0.239  \\
          & HSE03 (Ref.~\citenum{07PRB-Stroppa}) & -2.012  & -1.913  & -0.099  \bigstrut[b]\\
    \hline
    \hline
    \end{tabular}%
  \label{tab:addlabel}%
\end{table}%

Table~\ref{tab:Rh_adsorb} summarizes the differences in adsorption energy between CO molecules adsorbed at the top and fcc sites on the Rh surface calculated using various methods.
Similarly to the Cu system, PBE fails to reproduce experimental observations, erroneously predicting the fcc site to be more stable than the top site.
In contrast, HSE06 consistently identifies the top configuration as the energetically favorable adsorption geometry, regardless of whether the calculations are based on PBE-optimized or self-consistently relaxed structures.
In PBE and HSE06 calculations, the adsorption energies align well with literature results.
For DeePKS models, both the DeePKS-Rh and DeePKS-Cu+Rh models achieve qualitative agreement with HSE06 when evaluated on PBE-optimized geometries, with a slightly larger discrepancy of 0.01 eV than in Cu systems.
Self-consistent structural relaxations introduce marginally increased deviations up to 0.012 eV, yet all DeePKS models still correctly preserve the preference of experiments and HSE06.
Notably, direct application of the DeePKS-Cu model to the Rh systems yields significant errors of -0.34 eV in top-site adsorption energy.
The discrepancies exceed 1 eV/atom in several adsorption energy components, as shown in Supplementary Table S3.
This failure arises from non-overlapping descriptor spaces between Cu and Rh systems, underscoring the necessity of mixed training, like for DeePKS-Cu+Rh, for cross-elemental applications.

\subsection{Efficiency of DeePKS}
The computational efficiency of DeePKS models is evaluated using the adsorption system of CO molecules at the top site of the Rh(111) surface as a case study.
SCF calculations are carried out with 56 CPU cores of the Intel(R) Xeon(R) Gold 6348 CPU at 2.60GHz. 
The hybrid functional HSE requires approximately $1\times10^{4}$ seconds to complete the full calculation, while PBE finishes in 250 seconds. 
The DeePKS-Rh model achieves comparable accuracy with HSE but only takes around $1.2\times10^{3}$ seconds, representing a remarkable improvement over HSE while being significantly faster.
It's only about 4 times slower than PBE but 12 times faster than HSE. 
This demonstrates that the DeePKS approach offers a favorable balance between accuracy and computational cost.

\subsection{Potential Energy Surface}

\begin{figure}
  \includegraphics[width=14cm]{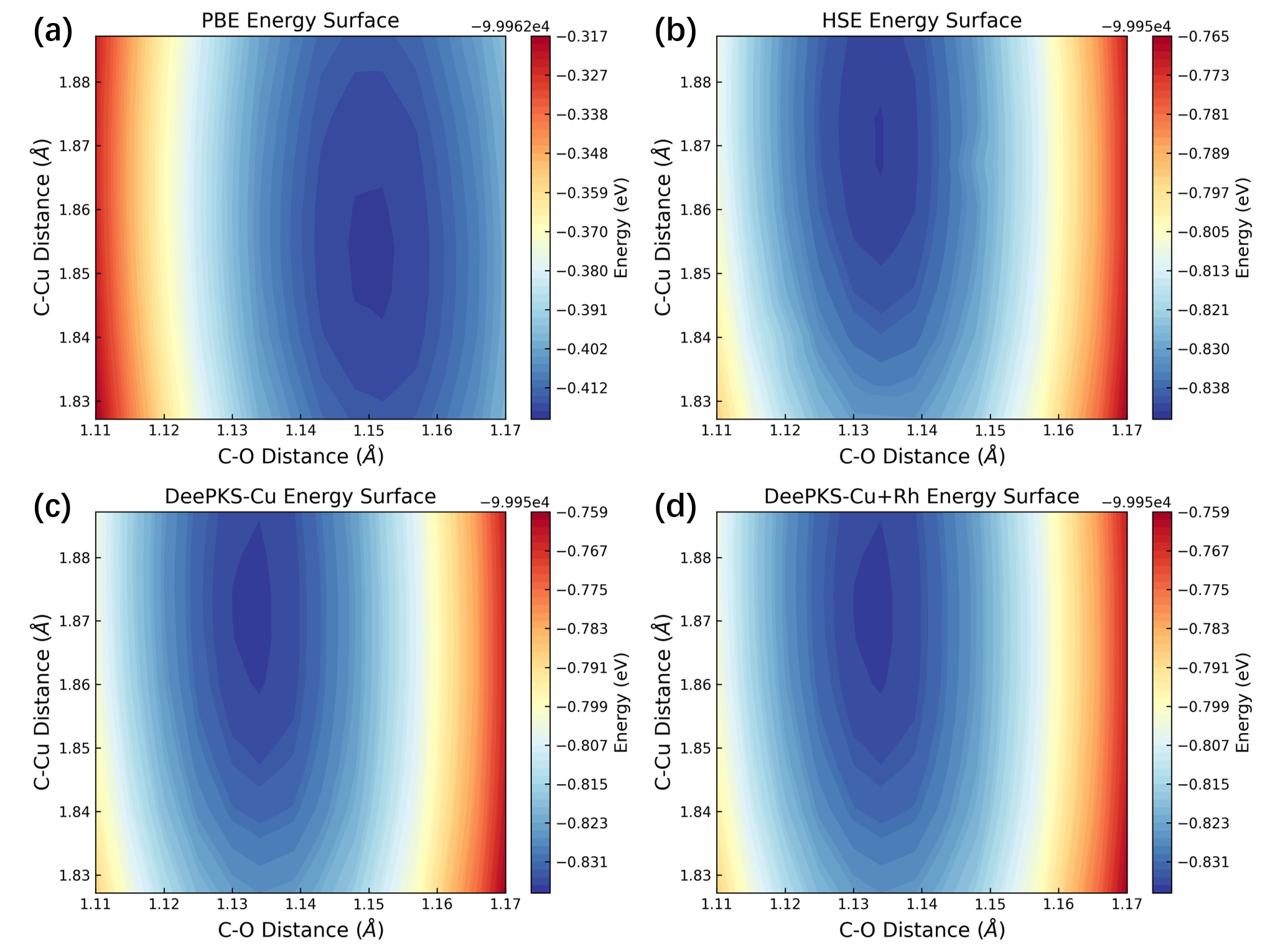}
  \caption{(Color online)
    Potential energy surface (PES) for CO adsorption at the top site of the Cu surface.
    The CO molecule maintains a vertical adsorption configuration.
    The horizontal axis represents the C-O bond distance (\AA) in the CO molecule, while the vertical axis denotes the distance between the C atom and the bonded Cu atom (\AA), with the remaining coordinates fixed at their PBE-optimized positions.
    Panels (a)-(d) display the PESs calculated by the PBE and HSE06 functionals, as well as the DeePKS-Cu and DeePKS-Cu+Rh models, respectively.
  }\label{pes}
\end{figure}

Fig.~\ref{pes} presents the potential energy surface (PES) of CO adsorbed at the top site of Cu(111) surface in a vertical adsorption configuration. 
The horizontal and vertical axes correspond to the C-O bond distance of CO molecule and the C-Cu distance between the C atom and the coordinating Cu atom, respectively, with the remaining coordinates fixed at their relaxed positions with the PBE functional. 
The PES was generated using 140 discrete DFT calculations. 
Comparative analysis reveals distinct minimum energy positions between PBE and HSE06-derived PESs.
In contrast, both DeePKS-Cu and DeePKS-Cu+Rh models exhibit PES minima closely aligned with the HSE06 results.
Furthermore, the DeePKS models reproduce HSE06's curvature characteristics across the PES, though a systematic vertical shift of 6 meV is observed. 
This energy discrepancy remains significantly smaller than the 10 eV differences between PBE and HSE06 across the sampled configuration space.

\begin{table}[htbp]
  \centering
  \caption{
  C-O and C-Cu bond distances (Å) in height for CO adsorbed at the top site of Cu surface under equilibrium configurations, calculated using different methods including the base functional PBE, target functional HSE06, DeePKS-Cu, and DeePKS-Cu+Rh models. 
  The C-Cu distance is defined as the distance between the carbon atom in CO and the coordinating Cu atom at the adsorption site.
  }
    \begin{tabular}{lcc}
    \hline
    \hline
    Method & $d_{\rm{C-O}}$ & $d_{\rm{C-Cu}}$ \\
    \hline
    PBE   & 1.150 & 1.850 \\
    PBE (Ref.~\citenum{07PRB-Stroppa}) & 1.158 & 1.844 \\
    HSE06   & 1.133  & 1.869  \\
    DeePKS-Cu & 1.133 & 1.868 \\
    DeePKS-Cu+Rh & 1.133 & 1.860 \\
    HSE03 (Ref.~\citenum{07PRB-Stroppa}) & 1.142 & 1.864 \\
    \hline
    \hline
    \end{tabular}%
  \label{tab:pes_distance}%
\end{table}%

To further verify the consistency of the PES minima, geometry relaxation is performed at the top site, and the optimized C-O and C-Cu bond distances are calculated.
Since the CO molecule maintains its vertical adsorption orientation after structural relaxation, the measured C-O and C-Cu distances here correspond directly to the height differences reported in Reference~\citenum{07PRB-Stroppa}.
%
The PBE and HSE06 relaxed geometries demonstrate excellent agreement with reference values, showing deviations lower than 0.01 \AA~in both bond distances. 
Impressively, the DeePKS models show remarkable accuracy in reproducing HSE06 reference data, achieving excellent agreement with bond length errors consistently below 0.001 \AA~across most configurations.
The DeePKS-Cu+Rh model shows slightly larger deviations of 0.009 \AA~in C-Cu bond lengths compared to HSE06 references, corresponding to a negligible $0.5\%$ difference. 
This error is still significantly smaller than the 0.02 \AA~variations observed between PBE and HSE06 methods, confirming the model’s accuracy for structural predictions.
%
%
The structural consistency between DeePKS and HSE06 geometries directly corroborates the PES analysis in Fig.~\ref{pes}, where both methods share nearly identical minimum energy positions.

\section{Conclusions}\label{sec:conclusions}

In conclusion, we demonstrate that the Deep Kohn-Sham (DeePKS) framework, as implemented in the ABACUS package, provides an effective machine learning solution in terms of electronic structure information to the long-standing CO adsorption puzzle across transition metal surfaces. 
By adopting the DeePKS models trained on two metal systems (Cu(111) and Rh(111) surfaces), we performed DFT calculations with self-consistent field method and successfully reproduced the accuracy of the hybrid functional HSE06 in determining both adsorption energies and relative site preferences, while maintaining the comparable computational efficiency of the base functional PBE. With about 10 meV energy difference in adsorption energies, the model consistently predicted top-site adsorption stabilization for CO on Cu(111) and Rh(111) surfaces, resolving the incorrect fcc-site preference inherent to conventional GGA functionals. 
Systematic analysis of potential energy surfaces and structural parameters further validated the physical consistency between the DeePKS and HSE06 methods.

By incorporating adsorption data from both Cu(111) and Rh(111) surfaces during training, the DeePKS-Cu+Rh model was developed. We found that the model successfully captures the key physical features of both substrate materials. This highlights the critical importance of incorporating multi-material training data to achieve cross-system transferability.
While single-metal models (DeePKS-Cu/Rh) exhibit failures in cross-system predictions, the DeePKS-Cu+Rh model systematically resolves these limitations, demonstrating superior compatibility across both metals while maintaining comparable accuracy to metal-specific models in their corresponding systems.
Principal component analysis confirmed that atomic descriptors naturally differentiate chemical environments across metals and adsorption configurations, suggesting a pathway for developing universal models.
This approach enables efficient exploration of multi-component catalytic systems.

In summary, we established a machine-learning-based exchange-correlation functional within the DFT framework for investigations of CO adsorption across Cu(111) and Rh(111) metallic surfaces.
We hope this work opens avenues for simulating complex surface reactions with hybrid-functional accuracy while circumventing prohibitive computational costs.
Currently, the DeepKS model yields CO frontier orbital energies closer to the base functional PBE rather than approaching HSE06 target values.
Since these orbitals play a significant role in adsorption site preference as described by the Blyholder model, further refinement may be needed to fully align with target electronic properties. 
Future efforts will resolve this through DeePKS-ES, which incorporates electronic structure properties such as energy levels.
Additionally, introducing physical constraints, like exact conditions and asymptotic behavior, could further improve transferability across diverse systems while reducing the training data requirements.

$\\$
{\bf Acknowledgements}
$\\$
The work of X.L, and M.C. was supported by the National Key R$\&$D Program of China under Grant No. 2025YFB3003603, the National Natural Science Foundation of China under Grant Nos.11988102 and 12135002, and the Fundamental Research Funds for the Central Universities, Peking University. The numerical simulations were performed on the High-Performance Computing Platform of CAPT.

\bibliography{reference}

\providecommand{\latin}[1]{#1}
\makeatletter
\providecommand{\doi}
  {\begingroup\let\do\@makeother\dospecials
  \catcode`\{=1 \catcode`\}=2 \doi@aux}
\providecommand{\doi@aux}[1]{\endgroup\texttt{#1}}
\makeatother
\providecommand*\mcitethebibliography{\thebibliography}
\csname @ifundefined\endcsname{endmcitethebibliography}  {\let\endmcitethebibliography\endthebibliography}{}
\begin{mcitethebibliography}{53}
\providecommand*\natexlab[1]{#1}
\providecommand*\mciteSetBstSublistMode[1]{}
\providecommand*\mciteSetBstMaxWidthForm[2]{}
\providecommand*\mciteBstWouldAddEndPuncttrue
  {\def\EndOfBibitem{\unskip.}}
\providecommand*\mciteBstWouldAddEndPunctfalse
  {\let\EndOfBibitem\relax}
\providecommand*\mciteSetBstMidEndSepPunct[3]{}
\providecommand*\mciteSetBstSublistLabelBeginEnd[3]{}
\providecommand*\EndOfBibitem{}
\mciteSetBstSublistMode{f}
\mciteSetBstMaxWidthForm{subitem}{(\alph{mcitesubitemcount})}
\mciteSetBstSublistLabelBeginEnd
  {\mcitemaxwidthsubitemform\space}
  {\relax}
  {\relax}

\bibitem[Hohenberg and Kohn(1964)Hohenberg, and Kohn]{64PR-Hohenberg}
Hohenberg,~P.; Kohn,~W. Inhomogeneous electron gas. \emph{Physical Review} \textbf{1964}, \emph{136}, B864\relax
\mciteBstWouldAddEndPuncttrue
\mciteSetBstMidEndSepPunct{\mcitedefaultmidpunct}
{\mcitedefaultendpunct}{\mcitedefaultseppunct}\relax
\EndOfBibitem
\bibitem[Kohn and Sham(1965)Kohn, and Sham]{65PR-Kohn}
Kohn,~W.; Sham,~L.~J. Self-consistent equations including exchange and correlation effects. \emph{Physical Review} \textbf{1965}, \emph{140}, A1133\relax
\mciteBstWouldAddEndPuncttrue
\mciteSetBstMidEndSepPunct{\mcitedefaultmidpunct}
{\mcitedefaultendpunct}{\mcitedefaultseppunct}\relax
\EndOfBibitem
\bibitem[Perdew(2001)]{01AIP-Perdew}
Perdew,~J.~P. Jacob's Ladder of Density Functional Approximations for the Exchange-Correlation Energy. {{AIP Conf. Proc.}} 2001; pp 1--20\relax
\mciteBstWouldAddEndPuncttrue
\mciteSetBstMidEndSepPunct{\mcitedefaultmidpunct}
{\mcitedefaultendpunct}{\mcitedefaultseppunct}\relax
\EndOfBibitem
\bibitem[Perdew \latin{et~al.}(1996)Perdew, Burke, and Ernzerhof]{96PRL-Perdew}
Perdew,~J.~P.; Burke,~K.; Ernzerhof,~M. Generalized Gradient Approximation Made Simple. \emph{Physical Review Letters} \textbf{1996}, \emph{77}, 3865--3868\relax
\mciteBstWouldAddEndPuncttrue
\mciteSetBstMidEndSepPunct{\mcitedefaultmidpunct}
{\mcitedefaultendpunct}{\mcitedefaultseppunct}\relax
\EndOfBibitem
\bibitem[Heyd \latin{et~al.}(2003)Heyd, Scuseria, and Ernzerhof]{03JCP-Heyd}
Heyd,~J.; Scuseria,~G.~E.; Ernzerhof,~M. Hybrid Functionals Based on a Screened {{Coulomb}} Potential. \emph{The Journal of Chemical Physics} \textbf{2003}, \emph{118}, 8207--8215\relax
\mciteBstWouldAddEndPuncttrue
\mciteSetBstMidEndSepPunct{\mcitedefaultmidpunct}
{\mcitedefaultendpunct}{\mcitedefaultseppunct}\relax
\EndOfBibitem
\bibitem[Krukau \latin{et~al.}(2006)Krukau, Vydrov, Izmaylov, and Scuseria]{06JCP-Krukau}
Krukau,~A.~V.; Vydrov,~O.~A.; Izmaylov,~A.~F.; Scuseria,~G.~E. Influence of the Exchange Screening Parameter on the Performance of Screened Hybrid Functionals. \emph{Journal of Chemical Physics} \textbf{2006}, \emph{125}, 224106\relax
\mciteBstWouldAddEndPuncttrue
\mciteSetBstMidEndSepPunct{\mcitedefaultmidpunct}
{\mcitedefaultendpunct}{\mcitedefaultseppunct}\relax
\EndOfBibitem
\bibitem[Perdew \latin{et~al.}(1996)Perdew, Ernzerhof, and Burke]{96JCP-Perdew}
Perdew,~J.~P.; Ernzerhof,~M.; Burke,~K. Rationale for Mixing Exact Exchange with Density Functional Approximations. \emph{Journal of Chemical Physics} \textbf{1996}, \emph{105}, 9982--9985\relax
\mciteBstWouldAddEndPuncttrue
\mciteSetBstMidEndSepPunct{\mcitedefaultmidpunct}
{\mcitedefaultendpunct}{\mcitedefaultseppunct}\relax
\EndOfBibitem
\bibitem[Feibelman \latin{et~al.}(2001)Feibelman, Hammer, N{\o}rskov, Wagner, Scheffler, Stumpf, Watwe, and Dumesic]{01JPCB-Feibelman}
Feibelman,~P.~J.; Hammer,~B.; N{\o}rskov,~J.~K.; Wagner,~F.; Scheffler,~M.; Stumpf,~R.; Watwe,~R.; Dumesic,~J. The {{CO}}/{{Pt}}(111) {{Puzzle}}. \emph{The Journal of Physical Chemistry B} \textbf{2001}, \emph{105}, 4018--4025\relax
\mciteBstWouldAddEndPuncttrue
\mciteSetBstMidEndSepPunct{\mcitedefaultmidpunct}
{\mcitedefaultendpunct}{\mcitedefaultseppunct}\relax
\EndOfBibitem
\bibitem[Kessler and Thieme(1977)Kessler, and Thieme]{77SS-Kessler}
Kessler,~J.; Thieme,~F. Chemisorption of {{Co}} on {{Differently Prepared Cu}}(111) {{Surfaces}}. \emph{Surface Science} \textbf{1977}, \emph{67}, 405--415\relax
\mciteBstWouldAddEndPuncttrue
\mciteSetBstMidEndSepPunct{\mcitedefaultmidpunct}
{\mcitedefaultendpunct}{\mcitedefaultseppunct}\relax
\EndOfBibitem
\bibitem[Steininger \latin{et~al.}(1982)Steininger, Lehwald, and Ibach]{82SS-Steininger}
Steininger,~H.; Lehwald,~S.; Ibach,~H. On the Adsorption of {{CO}} on {{Pt}}(111). \emph{Surface Science} \textbf{1982}, \emph{123}, 264--282\relax
\mciteBstWouldAddEndPuncttrue
\mciteSetBstMidEndSepPunct{\mcitedefaultmidpunct}
{\mcitedefaultendpunct}{\mcitedefaultseppunct}\relax
\EndOfBibitem
\bibitem[Ogletree \latin{et~al.}(1986)Ogletree, Van~Hove, and Somorjai]{86SS-Ogletree}
Ogletree,~D.~F.; Van~Hove,~M.~A.; Somorjai,~G.~A. {{LEED}} Intensity Analysis of the Structures of Clean {{Pt}}(111) and of {{CO}} Adsorbed on {{Pt}}(111) in the c(4 {\texttimes} 2) Arrangement. \emph{Surface Science} \textbf{1986}, \emph{173}, 351--365\relax
\mciteBstWouldAddEndPuncttrue
\mciteSetBstMidEndSepPunct{\mcitedefaultmidpunct}
{\mcitedefaultendpunct}{\mcitedefaultseppunct}\relax
\EndOfBibitem
\bibitem[Raval \latin{et~al.}(1988)Raval, Parker, Pemble, Hollins, Pritchard, and Chesters]{88SS-Raval}
Raval,~R.; Parker,~S.~F.; Pemble,~M.~E.; Hollins,~P.; Pritchard,~J.; Chesters,~M.~A. {{FT-rairs}}, Eels and Leed Studies of the Adsorption of Carbon Monoxide on {{Cu}}(111). \emph{Surface Science} \textbf{1988}, \emph{203}, 353--377\relax
\mciteBstWouldAddEndPuncttrue
\mciteSetBstMidEndSepPunct{\mcitedefaultmidpunct}
{\mcitedefaultendpunct}{\mcitedefaultseppunct}\relax
\EndOfBibitem
\bibitem[Grinberg \latin{et~al.}(2002)Grinberg, Yourdshahyan, and Rappe]{02JCP-Grinberg}
Grinberg,~I.; Yourdshahyan,~Y.; Rappe,~A.~M. {{CO}} on {{Pt}}(111) Puzzle: {{A}} Possible Solution. \emph{The Journal of Chemical Physics} \textbf{2002}, \emph{117}, 2264--2270\relax
\mciteBstWouldAddEndPuncttrue
\mciteSetBstMidEndSepPunct{\mcitedefaultmidpunct}
{\mcitedefaultendpunct}{\mcitedefaultseppunct}\relax
\EndOfBibitem
\bibitem[Gil(2003)]{03SurSci-Gil}
Gil,~A. Site Preference of {{CO}} Chemisorbed on {{Pt}}(111) from Density Functional Calculations. \emph{Surface Science} \textbf{2003}, \emph{530}, 71--87\relax
\mciteBstWouldAddEndPuncttrue
\mciteSetBstMidEndSepPunct{\mcitedefaultmidpunct}
{\mcitedefaultendpunct}{\mcitedefaultseppunct}\relax
\EndOfBibitem
\bibitem[Patra \latin{et~al.}(2019)Patra, Peng, Sun, and Perdew]{19PRB-Patra}
Patra,~A.; Peng,~H.; Sun,~J.; Perdew,~J.~P. Rethinking {{CO}} Adsorption on Transition-Metal Surfaces: {{Effect}} of Density-Driven Self-Interaction Errors. \emph{Physical Review B} \textbf{2019}, \emph{100}, 035442\relax
\mciteBstWouldAddEndPuncttrue
\mciteSetBstMidEndSepPunct{\mcitedefaultmidpunct}
{\mcitedefaultendpunct}{\mcitedefaultseppunct}\relax
\EndOfBibitem
\bibitem[Blyholder(1964)]{64JPC-Blyholder}
Blyholder,~G. Molecular Orbital View of Chemisorbed Carbon Monoxide. \emph{Journal of Physical Chemistry} \textbf{1964}, \emph{68}, 2772--2777\relax
\mciteBstWouldAddEndPuncttrue
\mciteSetBstMidEndSepPunct{\mcitedefaultmidpunct}
{\mcitedefaultendpunct}{\mcitedefaultseppunct}\relax
\EndOfBibitem
\bibitem[Aizawa and Tsuneyuki(1998)Aizawa, and Tsuneyuki]{98SurSci-Aizawa}
Aizawa,~H.; Tsuneyuki,~S. First-Principles Study of {{CO}} Bonding to {{Pt}}(111): Validity of the {{Blyholder}} Model. \emph{Surface Science} \textbf{1998}, \emph{399}, L364--L370\relax
\mciteBstWouldAddEndPuncttrue
\mciteSetBstMidEndSepPunct{\mcitedefaultmidpunct}
{\mcitedefaultendpunct}{\mcitedefaultseppunct}\relax
\EndOfBibitem
\bibitem[Illas \latin{et~al.}(1995)Illas, Zurita, Rubio, and M{\'a}rquez]{95PRB-Illas}
Illas,~F.; Zurita,~S.; Rubio,~J.; M{\'a}rquez,~A.~M. Origin of the Vibrational Shift of {{CO}} Chemisorbed on {{Pt}}(111). \emph{Physical Review B} \textbf{1995}, \emph{52}, 12372--12379\relax
\mciteBstWouldAddEndPuncttrue
\mciteSetBstMidEndSepPunct{\mcitedefaultmidpunct}
{\mcitedefaultendpunct}{\mcitedefaultseppunct}\relax
\EndOfBibitem
\bibitem[Koper \latin{et~al.}(2000)Koper, Van~Santen, Wasileski, and Weaver]{00JCP-Koper}
Koper,~M. T.~M.; Van~Santen,~R.~A.; Wasileski,~S.~A.; Weaver,~M.~J. Field-Dependent Chemisorption of Carbon Monoxide and Nitric Oxide on Platinum-Group (111) Surfaces: Quantum Chemical Calculations Compared with Infrared Spectroscopy at Electrochemical and Vacuum-Based Interfaces. \emph{Journal of Chemical Physics} \textbf{2000}, \emph{113}, 4392--4407\relax
\mciteBstWouldAddEndPuncttrue
\mciteSetBstMidEndSepPunct{\mcitedefaultmidpunct}
{\mcitedefaultendpunct}{\mcitedefaultseppunct}\relax
\EndOfBibitem
\bibitem[Doll(2004)]{04SurSci-Doll}
Doll,~K. {{CO}} Adsorption on the {{Pt}}(111) Surface: A Comparison of a Gradient Corrected Functional and a Hybrid Functional. \emph{Surface Science} \textbf{2004}, \emph{573}, 464--473\relax
\mciteBstWouldAddEndPuncttrue
\mciteSetBstMidEndSepPunct{\mcitedefaultmidpunct}
{\mcitedefaultendpunct}{\mcitedefaultseppunct}\relax
\EndOfBibitem
\bibitem[Gameel \latin{et~al.}(2018)Gameel, Sharafeldin, Abourayya, Biby, and Allam]{18PCCP-Gameel}
Gameel,~K.~M.; Sharafeldin,~I.~M.; Abourayya,~A.~U.; Biby,~A.~H.; Allam,~N.~K. Unveiling {{CO}} Adsorption on {{Cu}} Surfaces: New Insights from Molecular Orbital Principles. \emph{Physical Chemistry Chemical Physics} \textbf{2018}, \emph{20}, 25892--25900\relax
\mciteBstWouldAddEndPuncttrue
\mciteSetBstMidEndSepPunct{\mcitedefaultmidpunct}
{\mcitedefaultendpunct}{\mcitedefaultseppunct}\relax
\EndOfBibitem
\bibitem[Lazi{\'c} \latin{et~al.}(2010)Lazi{\'c}, Alaei, Atodiresei, Caciuc, Brako, and Bl{\"u}gel]{10PRB-Lazic}
Lazi{\'c},~P.; Alaei,~M.; Atodiresei,~N.; Caciuc,~V.; Brako,~R.; Bl{\"u}gel,~S. Density Functional Theory with Nonlocal Correlation: A Key to the Solution of the {{CO}} Adsorption Puzzle. \emph{Physical Review B} \textbf{2010}, \emph{81}, 45401\relax
\mciteBstWouldAddEndPuncttrue
\mciteSetBstMidEndSepPunct{\mcitedefaultmidpunct}
{\mcitedefaultendpunct}{\mcitedefaultseppunct}\relax
\EndOfBibitem
\bibitem[Kresse \latin{et~al.}(2003)Kresse, Gil, and Sautet]{03PRB-Kresse}
Kresse,~G.; Gil,~A.; Sautet,~P. Significance of Single-Electron Energies for the Description of {{CO}} on {{Pt}}(111). \emph{Physical Review B} \textbf{2003}, \emph{68}, 073401\relax
\mciteBstWouldAddEndPuncttrue
\mciteSetBstMidEndSepPunct{\mcitedefaultmidpunct}
{\mcitedefaultendpunct}{\mcitedefaultseppunct}\relax
\EndOfBibitem
\bibitem[K{\"o}hler and Kresse(2004)K{\"o}hler, and Kresse]{04PRB-Kohler}
K{\"o}hler,~L.; Kresse,~G. Density Functional Study of {{CO}} on {{Rh}}(111). \emph{Physical Review B} \textbf{2004}, \emph{70}, 165405\relax
\mciteBstWouldAddEndPuncttrue
\mciteSetBstMidEndSepPunct{\mcitedefaultmidpunct}
{\mcitedefaultendpunct}{\mcitedefaultseppunct}\relax
\EndOfBibitem
\bibitem[Gajdo{\v s} and Hafner(2005)Gajdo{\v s}, and Hafner]{05SurSci-Gajdos}
Gajdo{\v s},~M.; Hafner,~J. {{CO}} Adsorption on {{Cu}}(111) and {{Cu}}(001) Surfaces: {{Improving}} Site Preference in {{DFT}} Calculations. \emph{Surface Science} \textbf{2005}, \emph{590}, 117--126\relax
\mciteBstWouldAddEndPuncttrue
\mciteSetBstMidEndSepPunct{\mcitedefaultmidpunct}
{\mcitedefaultendpunct}{\mcitedefaultseppunct}\relax
\EndOfBibitem
\bibitem[Hammer \latin{et~al.}(1999)Hammer, Hansen, and N{\o}rskov]{99PRB-Hammer}
Hammer,~B.; Hansen,~L.~B.; N{\o}rskov,~J.~K. Improved Adsorption Energetics within Density-Functional Theory Using Revised {{Perdew-Burke-Ernzerhof}} Functionals. \emph{Physical Review B} \textbf{1999}, \emph{59}, 7413--7421\relax
\mciteBstWouldAddEndPuncttrue
\mciteSetBstMidEndSepPunct{\mcitedefaultmidpunct}
{\mcitedefaultendpunct}{\mcitedefaultseppunct}\relax
\EndOfBibitem
\bibitem[Gajdo \latin{et~al.}(2004)Gajdo, Eichler, and Hafner]{04JPCM-Gajdo}
Gajdo,~M.; Eichler,~A.; Hafner,~J. {{CO}} Adsorption on Close-Packed Transition and Noble Metal Surfaces: Trends from {\emph{Ab Initio}} Calculations. \emph{Journal of Physics: Condensed Matter} \textbf{2004}, \emph{16}, 1141--1164\relax
\mciteBstWouldAddEndPuncttrue
\mciteSetBstMidEndSepPunct{\mcitedefaultmidpunct}
{\mcitedefaultendpunct}{\mcitedefaultseppunct}\relax
\EndOfBibitem
\bibitem[Wellendorff \latin{et~al.}(2012)Wellendorff, Lundgaard, Møgelhøj, Petzold, Landis, Nørskov, Bligaard, and Jacobsen]{12PRB-Wellendorff}
Wellendorff,~J.; Lundgaard,~K.~T.; Møgelhøj,~A.; Petzold,~V.; Landis,~D.~D.; Nørskov,~J.~K.; Bligaard,~T.; Jacobsen,~K.~W. Density functionals for surface science: Exchange-correlation model development with Bayesian error estimation. \emph{Physical Review B} \textbf{2012}, \emph{85}\relax
\mciteBstWouldAddEndPuncttrue
\mciteSetBstMidEndSepPunct{\mcitedefaultmidpunct}
{\mcitedefaultendpunct}{\mcitedefaultseppunct}\relax
\EndOfBibitem
\bibitem[Lundgaard \latin{et~al.}(2016)Lundgaard, Wellendorff, Voss, Jacobsen, and Bligaard]{16PRB-Lundgaard}
Lundgaard,~K.~T.; Wellendorff,~J.; Voss,~J.; Jacobsen,~K.~W.; Bligaard,~T. mBEEF-vdW: Robust fitting of error estimation density functionals. \emph{Physical Review B} \textbf{2016}, \emph{93}\relax
\mciteBstWouldAddEndPuncttrue
\mciteSetBstMidEndSepPunct{\mcitedefaultmidpunct}
{\mcitedefaultendpunct}{\mcitedefaultseppunct}\relax
\EndOfBibitem
\bibitem[Wellendorff \latin{et~al.}()Wellendorff, Silbaugh, Garcia-Pintos, Nørskov, Bligaard, Studt, and Campbell]{15SS-Wellendorff}
Wellendorff,~J.; Silbaugh,~T.~L.; Garcia-Pintos,~D.; Nørskov,~J.~K.; Bligaard,~T.; Studt,~F.; Campbell,~C.~T. A Benchmark Database for Adsorption Bond Energies to Transition Metal Surfaces and Comparison to Selected {{DFT}} Functionals. \emph{640}, 36--44\relax
\mciteBstWouldAddEndPuncttrue
\mciteSetBstMidEndSepPunct{\mcitedefaultmidpunct}
{\mcitedefaultendpunct}{\mcitedefaultseppunct}\relax
\EndOfBibitem
\bibitem[Araujo \latin{et~al.}(2022)Araujo, Rodrigues, Dos~Santos, and Pettersson]{22NC-Araujo}
Araujo,~R.~B.; Rodrigues,~G. L.~S.; Dos~Santos,~E.~C.; Pettersson,~L. G.~M. Adsorption Energies on Transition Metal Surfaces: Towards an Accurate and Balanced Description. \emph{Nature Communications} \textbf{2022}, \emph{13}, 6853\relax
\mciteBstWouldAddEndPuncttrue
\mciteSetBstMidEndSepPunct{\mcitedefaultmidpunct}
{\mcitedefaultendpunct}{\mcitedefaultseppunct}\relax
\EndOfBibitem
\bibitem[Stroppa \latin{et~al.}(2007)Stroppa, Termentzidis, Paier, Kresse, and Hafner]{07PRB-Stroppa}
Stroppa,~A.; Termentzidis,~K.; Paier,~J.; Kresse,~G.; Hafner,~J. {{CO}} Adsorption on Metal Surfaces: {{A}} Hybrid Functional Study with Plane-Wave Basis Set. \emph{Physical Review B} \textbf{2007}, \emph{76}, 195440\relax
\mciteBstWouldAddEndPuncttrue
\mciteSetBstMidEndSepPunct{\mcitedefaultmidpunct}
{\mcitedefaultendpunct}{\mcitedefaultseppunct}\relax
\EndOfBibitem
\bibitem[Stroppa and Kresse(2008)Stroppa, and Kresse]{08NJP-Stroppa}
Stroppa,~A.; Kresse,~G. The Shortcomings of Semi-Local and Hybrid Functionals: What We Can Learn from Surface Science Studies. \emph{New Journal of Physics} \textbf{2008}, \emph{10}, 063020\relax
\mciteBstWouldAddEndPuncttrue
\mciteSetBstMidEndSepPunct{\mcitedefaultmidpunct}
{\mcitedefaultendpunct}{\mcitedefaultseppunct}\relax
\EndOfBibitem
\bibitem[Ren \latin{et~al.}(2012)Ren, Rinke, Joas, and Scheffler]{12JMS-Ren}
Ren,~X.; Rinke,~P.; Joas,~C.; Scheffler,~M. Random-phase approximation and its applications in computational chemistry and materials science. \emph{Journal of Materials Science} \textbf{2012}, \emph{47}, 7447--7471\relax
\mciteBstWouldAddEndPuncttrue
\mciteSetBstMidEndSepPunct{\mcitedefaultmidpunct}
{\mcitedefaultendpunct}{\mcitedefaultseppunct}\relax
\EndOfBibitem
\bibitem[Ren \latin{et~al.}(2009)Ren, Rinke, and Scheffler]{09PRB-Ren}
Ren,~X.; Rinke,~P.; Scheffler,~M. Exploring the Random Phase Approximation: {{Application}} to {{CO}} Adsorbed on {{Cu}}(111). \emph{Physical Review B} \textbf{2009}, \emph{80}, 045402\relax
\mciteBstWouldAddEndPuncttrue
\mciteSetBstMidEndSepPunct{\mcitedefaultmidpunct}
{\mcitedefaultendpunct}{\mcitedefaultseppunct}\relax
\EndOfBibitem
\bibitem[Schimka \latin{et~al.}(2010)Schimka, Harl, Stroppa, Gr{\"u}neis, Marsman, Mittendorfer, and Kresse]{10NM-Schimka}
Schimka,~L.; Harl,~J.; Stroppa,~A.; Gr{\"u}neis,~A.; Marsman,~M.; Mittendorfer,~F.; Kresse,~G. Accurate Surface and Adsorption Energies from Many-Body Perturbation Theory. \emph{Nature Materials} \textbf{2010}, \emph{9}, 741--744\relax
\mciteBstWouldAddEndPuncttrue
\mciteSetBstMidEndSepPunct{\mcitedefaultmidpunct}
{\mcitedefaultendpunct}{\mcitedefaultseppunct}\relax
\EndOfBibitem
\bibitem[Batatia \latin{et~al.}()Batatia, Benner, Chiang, Elena, Kovács, Riebesell, Advincula, Asta, Baldwin, Bernstein, Bhowmik, Blau, Cărare, Darby, De, Pia, Deringer, Elijošius, El-Machachi, Fako, Ferrari, Genreith-Schriever, George, Goodall, Grey, Han, Handley, Heenen, Hermansson, Holm, Jaafar, Hofmann, Jakob, Jung, Kapil, Kaplan, Karimitari, Kroupa, Kullgren, Kuner, Kuryla, Liepuoniute, Margraf, Magdău, Michaelides, Moore, Naik, Niblett, Norwood, O'Neill, Ortner, Persson, Reuter, Rosen, Schaaf, Schran, Sivonxay, Stenczel, Svahn, Sutton, family=Oord, Varga-Umbrich, Vegge, Vondrák, Wang, Witt, Zills, and Csányi]{23-Batatia}
Batatia,~I.; Benner,~P.; Chiang,~Y.; Elena,~A.~M.; Kovács,~D.~P.; Riebesell,~J.; Advincula,~X.~R.; Asta,~M.; Baldwin,~W.~J.; Bernstein,~N.; Bhowmik,~A.; Blau,~S.~M.; Cărare,~V.; Darby,~J.~P.; De,~S.; Pia,~F.~D.; Deringer,~V.~L.; Elijošius,~R.; El-Machachi,~Z.; Fako,~E.; Ferrari,~A.~C.; Genreith-Schriever,~A.; George,~J.; Goodall,~R. E.~A.; Grey,~C.~P.; Han,~S.; Handley,~W.; Heenen,~H.~H.; Hermansson,~K.; Holm,~C.; Jaafar,~J.; Hofmann,~S.; Jakob,~K.~S.; Jung,~H.; Kapil,~V.; Kaplan,~A.~D.; Karimitari,~N.; Kroupa,~N.; Kullgren,~J.; Kuner,~M.~C.; Kuryla,~D.; Liepuoniute,~G.; Margraf,~J.~T.; Magdău,~I.-B.; Michaelides,~A.; Moore,~J.~H.; Naik,~A.~A.; Niblett,~S.~P.; Norwood,~S.~W.; O'Neill,~N.; Ortner,~C.; Persson,~K.~A.; Reuter,~K.; Rosen,~A.~S.; Schaaf,~L.~L.; Schran,~C.; Sivonxay,~E.; Stenczel,~T.~K.; Svahn,~V.; Sutton,~C.; family=Oord,~p. d.~u.,~given=Cas; Varga-Umbrich,~E.; Vegge,~T.; Vondrák,~M.; Wang,~Y.; Witt,~W.~C.; Zills,~F.; Csányi,~G. A Foundation Model for Atomistic Materials Chemistry.
  \url{http://arxiv.org/abs/2401.00096}\relax
\mciteBstWouldAddEndPuncttrue
\mciteSetBstMidEndSepPunct{\mcitedefaultmidpunct}
{\mcitedefaultendpunct}{\mcitedefaultseppunct}\relax
\EndOfBibitem
\bibitem[Liu \latin{et~al.}(2023)Liu, Wang, Avargues, Verdi, Singraber, Karsai, Chen, and Kresse]{23PRL-Liu}
Liu,~P.; Wang,~J.; Avargues,~N.; Verdi,~C.; Singraber,~A.; Karsai,~F.; Chen,~X.-Q.; Kresse,~G. Combining {{Machine Learning}} and {{Many-Body Calculations}}: {{Coverage-Dependent Adsorption}} of {{CO}} on {{Rh}}(111). \emph{Physical Review Letters} \textbf{2023}, \emph{130}, 078001\relax
\mciteBstWouldAddEndPuncttrue
\mciteSetBstMidEndSepPunct{\mcitedefaultmidpunct}
{\mcitedefaultendpunct}{\mcitedefaultseppunct}\relax
\EndOfBibitem
\bibitem[Wu \latin{et~al.}(2025)Wu, Zheng, Zhang, Zhang, Li, and Pan]{25JMI-Wu}
Wu,~S.; Zheng,~S.; Zhang,~W.; Zhang,~M.; Li,~S.; Pan,~F. Machine-Learning Prediction of Facet-Dependent {{CO}} Coverage on {{Cu}} Electrocatalysts. \emph{Journal of Materials Informatics} \textbf{2025}, \emph{5}, N/A--N/A\relax
\mciteBstWouldAddEndPuncttrue
\mciteSetBstMidEndSepPunct{\mcitedefaultmidpunct}
{\mcitedefaultendpunct}{\mcitedefaultseppunct}\relax
\EndOfBibitem
\bibitem[S.~Muzas \latin{et~al.}(2024)S.~Muzas, Serrano~Jim{\'e}nez, Zhang, Jiang, Juaristi, and Alducin]{24JPCL-Muzas}
S.~Muzas,~A.; Serrano~Jim{\'e}nez,~A.; Zhang,~Y.; Jiang,~B.; Juaristi,~J.~I.; Alducin,~M. Multicoverage Study of Femtosecond Laser-Induced Desorption of {{CO}} from Pd(111). \emph{Journal of Physical Chemistry Letters} \textbf{2024}, \emph{15}, 2587--2594\relax
\mciteBstWouldAddEndPuncttrue
\mciteSetBstMidEndSepPunct{\mcitedefaultmidpunct}
{\mcitedefaultendpunct}{\mcitedefaultseppunct}\relax
\EndOfBibitem
\bibitem[{\v Z}ugec \latin{et~al.}(2024){\v Z}ugec, Tetenoire, Muzas, Zhang, Jiang, Alducin, and Juaristi]{24JACSAu-Zugec}
{\v Z}ugec,~I.; Tetenoire,~A.; Muzas,~A.~S.; Zhang,~Y.; Jiang,~B.; Alducin,~M.; Juaristi,~J.~I. Understanding the Photoinduced Desorption and Oxidation of {{CO}} on {{Ru}}(0001) Using a Neural Network Potential Energy Surface. \emph{JACS Au} \textbf{2024}, \emph{4}, 1997--2004\relax
\mciteBstWouldAddEndPuncttrue
\mciteSetBstMidEndSepPunct{\mcitedefaultmidpunct}
{\mcitedefaultendpunct}{\mcitedefaultseppunct}\relax
\EndOfBibitem
\bibitem[Guo \latin{et~al.}(2024)Guo, Song, and Liu]{24CS-Guo}
Guo,~Z.-X.; Song,~G.-L.; Liu,~Z.-P. Artificial Intelligence Driven Molecule Adsorption Prediction ({{AIMAP}}) Applied to Chirality Recognition of Amino Acid Adsorption on Metals. \emph{Chemical Science} \textbf{2024}, \emph{15}, 13369--13380\relax
\mciteBstWouldAddEndPuncttrue
\mciteSetBstMidEndSepPunct{\mcitedefaultmidpunct}
{\mcitedefaultendpunct}{\mcitedefaultseppunct}\relax
\EndOfBibitem
\bibitem[Agarwal and Joshi(2022)Agarwal, and Joshi]{22C-Agarwal}
Agarwal,~S.; Joshi,~K. Looking beyond Adsorption Energies to Understand Interactions at Surface Using Machine Learning. \emph{Chemistryselect} \textbf{2022}, \emph{7}, e202202414\relax
\mciteBstWouldAddEndPuncttrue
\mciteSetBstMidEndSepPunct{\mcitedefaultmidpunct}
{\mcitedefaultendpunct}{\mcitedefaultseppunct}\relax
\EndOfBibitem
\bibitem[Chen \latin{et~al.}(2021)Chen, Zhang, Wang, and E]{21JCTC-Chen}
Chen,~Y.; Zhang,~L.; Wang,~H.; E,~W. {{DeePKS}}: {{A Comprehensive Data-Driven Approach}} toward {{Chemically Accurate Density Functional Theory}}. \emph{J. Chem. Theory Comput.} \textbf{2021}, \emph{17}, 170--181\relax
\mciteBstWouldAddEndPuncttrue
\mciteSetBstMidEndSepPunct{\mcitedefaultmidpunct}
{\mcitedefaultendpunct}{\mcitedefaultseppunct}\relax
\EndOfBibitem
\bibitem[Li \latin{et~al.}(2022)Li, Ou, Chen, Cao, Liu, Zhang, Zheng, Cai, Wu, Wang, Chen, and Zhang]{22JPCA-Li}
Li,~W.; Ou,~Q.; Chen,~Y.; Cao,~Y.; Liu,~R.; Zhang,~C.; Zheng,~D.; Cai,~C.; Wu,~X.; Wang,~H.; Chen,~M.; Zhang,~L. {{DeePKS}}+{{ABACUS}} as a {{Bridge}} between {{Expensive Quantum Mechanical Models}} and {{Machine Learning Potentials}}. \emph{J. Phys. Chem. A} \textbf{2022}, \emph{126}, 9154--9164\relax
\mciteBstWouldAddEndPuncttrue
\mciteSetBstMidEndSepPunct{\mcitedefaultmidpunct}
{\mcitedefaultendpunct}{\mcitedefaultseppunct}\relax
\EndOfBibitem
\bibitem[Ou \latin{et~al.}(2023)Ou, Tuo, Li, Wang, Chen, and Zhang]{23JPCC-Ou}
Ou,~Q.; Tuo,~P.; Li,~W.; Wang,~X.; Chen,~Y.; Zhang,~L. {{DeePKS Model}} for {{Halide Perovskites}} with the {{Accuracy}} of a {{Hybrid Functional}}. \emph{J. Phys. Chem. C} \textbf{2023}, \emph{127}, 18755--18764\relax
\mciteBstWouldAddEndPuncttrue
\mciteSetBstMidEndSepPunct{\mcitedefaultmidpunct}
{\mcitedefaultendpunct}{\mcitedefaultseppunct}\relax
\EndOfBibitem
\bibitem[Liang \latin{et~al.}()Liang, Liu, and Chen]{25JCTC-Liang}
Liang,~X.; Liu,~R.; Chen,~M. A Deep Learning Framework for the Electronic Structure of Water: Toward a Universal Model. \emph{Journal of Chemical Theory and Computation} \relax
\mciteBstWouldAddEndPunctfalse
\mciteSetBstMidEndSepPunct{\mcitedefaultmidpunct}
{}{\mcitedefaultseppunct}\relax
\EndOfBibitem
\bibitem[Chen \latin{et~al.}(2010)Chen, Guo, and He]{10JPCM-Chen}
Chen,~M.; Guo,~G.-C.; He,~L. Systematically Improvable Optimized Atomic Basis Sets for Ab Initio Calculations. \emph{Journal of Physics: Condensed Matter} \textbf{2010}, \emph{22}, 445501\relax
\mciteBstWouldAddEndPuncttrue
\mciteSetBstMidEndSepPunct{\mcitedefaultmidpunct}
{\mcitedefaultendpunct}{\mcitedefaultseppunct}\relax
\EndOfBibitem
\bibitem[Li \latin{et~al.}(2016)Li, Liu, Chen, Lin, Ren, Lin, Yang, and He]{16CMS-Li}
Li,~P.; Liu,~X.; Chen,~M.; Lin,~P.; Ren,~X.; Lin,~L.; Yang,~C.; He,~L. Large-Scale {\emph{Ab Initio}} Simulations Based on Systematically Improvable Atomic Basis. \emph{Computational Materials Science} \textbf{2016}, \emph{112}, 503--517\relax
\mciteBstWouldAddEndPuncttrue
\mciteSetBstMidEndSepPunct{\mcitedefaultmidpunct}
{\mcitedefaultendpunct}{\mcitedefaultseppunct}\relax
\EndOfBibitem
\bibitem[Schlipf and Gygi(2015)Schlipf, and Gygi]{15CPC-Schlipf}
Schlipf,~M.; Gygi,~F. Optimization Algorithm for the Generation of {{ONCV}} Pseudopotentials. \emph{Computer Physics Communications} \textbf{2015}, \emph{196}, 36--44\relax
\mciteBstWouldAddEndPuncttrue
\mciteSetBstMidEndSepPunct{\mcitedefaultmidpunct}
{\mcitedefaultendpunct}{\mcitedefaultseppunct}\relax
\EndOfBibitem
\bibitem[Johnson(1988)]{88PRB-Johnson}
Johnson,~D.~D. Modified Broyden's method for accelerating convergence in self-consistent calculations. \emph{Physical Review B} \textbf{1988}, \emph{38}, 12807--12813\relax
\mciteBstWouldAddEndPuncttrue
\mciteSetBstMidEndSepPunct{\mcitedefaultmidpunct}
{\mcitedefaultendpunct}{\mcitedefaultseppunct}\relax
\EndOfBibitem
\bibitem[Huber and Herzberg()Huber, and Herzberg]{79-Huber}
Huber,~K.~P.; Herzberg,~G. \emph{Molecular {{Spectra}} and {{Molecular Structure}}: {{IV}}. {{Constants}} of {{Diatomic Molecules}}}; Springer US, pp 8--689\relax
\mciteBstWouldAddEndPuncttrue
\mciteSetBstMidEndSepPunct{\mcitedefaultmidpunct}
{\mcitedefaultendpunct}{\mcitedefaultseppunct}\relax
\EndOfBibitem
\end{mcitethebibliography}

\end{document}


\title{Supplemental Material for ``Investigating CO Adsorption on  Cu(111) and Rh(111) Surfaces Using Machine Learning Exchange-Correlation Functionals"}

\author{Xinyuan Liang}
\affiliation{Academy for Advanced Interdisciplinary Studies, Peking University, Beijing, 90871, P. R. China}
\affiliation{HEDPS, CAPT, College of Engineering, Peking University, Beijing, 100871, P. R. China}
\author{Renxi Liu}
\affiliation{Academy for Advanced Interdisciplinary Studies, Peking University, Beijing, 90871, P. R. China}
\affiliation{HEDPS, CAPT, College of Engineering, Peking University, Beijing, 100871, P. R. China}
\author{Mohan Chen}
\thanks{$^\ast$Corresponding author. Email:mohanchen@pku.edu.cn}
\affiliation{Academy for Advanced Interdisciplinary Studies, Peking University, Beijing, 90871, P. R. China}
\affiliation{HEDPS, CAPT, College of Engineering, Peking University, Beijing, 100871, P. R. China}

\maketitle

\section{Number of Data in the Dataset}

\begin{table}[htbp]
  \centering
  \caption{
  Number of data in the dataset, including configurations of isolated CO molecules, bare Cu(111) and Rh(111) surfaces, and CO-adsorbed surfaces at top, fcc, hcp, and bridge sites of both metals.
  }
    \begin{tabular}{lllccc}
    \hline
    \hline
    \multicolumn{3}{l}{System} & Number of data && Total number of data \bigstrut\\
    \hline
    \multicolumn{3}{l}{CO} & 10    && 10 \bigstrut\\
    \hline
    \multirow{5}[2]{*}{Cu} && bare surface & 10    && \multirow{5}[1]{*}{90} \bigstrut[t]\\
          && adsorbed surface: top & 20    &&  \\
          && adsorbed surface: fcc & 20    &&  \\
          && adsorbed surface: hcp & 20    &&  \\
          && adsorbed surface: bridge & 20    &&  \bigstrut[b]\\
    \hline
    \multirow{5}[1]{*}{Rh} && bare surface & 20    && \multirow{5}[1]{*}{100} \bigstrut[t]\\
          && adsorbed surface: top & 20    &&  \\
          && adsorbed surface: fcc & 20    &&  \\
          && adsorbed surface: hcp & 20    &&  \\
          && adsorbed surface: bridge & 20    &&  \\
    \hline
    \hline
    \end{tabular}%
  \label{tab:addlabel}%
\end{table}%

\newpage
\section{Energies related to Cu Adsorption System}

\begin{table}[htbp]
  \centering
  \caption{
  Energy components (eV) relevant to adsorption energy calculations in Cu systems, including gas-phase CO molecule energy, bare surface energy, and energies of top-site and fcc-site adsorption systems, calculated using different computational methods: base functional PBE, target functional HSE, and several DeePKS models. 
  For HSE and DeePKS methods, adsorption energies are evaluated under two geometric conditions: (i) PBE-optimized configurations and (ii) self-relaxed geometries (method-specific structural optimizations).
  }
    \begin{tabular}{lcccc}
    \hline
    \hline
    Method & CO    & Bare surface & Top   & Fcc \bigstrut\\
    \hline
    PBE   & -588.354  & -198746.648  & -199924.838  & -199925.057  \bigstrut[t]\\
    \multicolumn{5}{c}{Energies calculated at PBE-relaxed geometries}\rule{0pt}{12pt} \\
    HSE   & -587.964  & -198724.570  & -199901.643  & -199901.472  \\
    DeePKS-Cu & -587.963  & -198724.547  & -199901.618  & -199901.451  \\
    DeePKS-Cu+Rh & -587.964  & -198724.555  & -199901.643  & -199901.466  \\
    \multicolumn{5}{c}{Energies calculated at method-relaxed geometries}\rule{0pt}{12pt} \\
    HSE   & -587.974  & -198724.570  & -199901.680  & -199901.531  \\
    DeePKS-Cu & -587.974  & -198724.551  & -199901.656  & -199901.511  \\
    DeePKS-Cu+Rh & -587.975  & -198724.562  & -199901.680  & -199901.528  \bigstrut[b]\\
    \hline
    \hline
    \end{tabular}%
  \label{tab:addlabel}%
\end{table}%

\newpage
\section{Energies related to Rh Adsorption System}

\begin{table}[htbp]
  \centering
  \caption{
  Energy components (eV) relevant to adsorption energy calculations in Rh systems, including gas-phase CO molecule energy, bare surface energy, and energies of top-site and fcc-site adsorption systems, calculated using different computational methods: base functional PBE, target functional HSE, and several DeePKS models. 
  For HSE and DeePKS methods, adsorption energies are evaluated under two geometric conditions: (i) PBE-optimized configurations and (ii) self-relaxed geometries (method-specific structural optimizations).
  }
    \begin{tabular}{lcccc}
    \hline
    \hline
    Method & CO    & Bare surface & Top   & Fcc \bigstrut\\
    \hline
    PBE   & -588.354  & -118113.422  & -119294.085  & -119294.095  \bigstrut[t]\\
    \multicolumn{5}{c}{Energies calculated at PBE-relaxed geometries}\rule{0pt}{12pt} \\
    HSE   & -587.964  & -118052.769  & -119233.076  & -119232.727  \\
    DeePKS-Rh & -587.965  & -118052.740  & -119233.059  & -119232.725  \\
    DeePKS-Cu+Rh & -587.963  & -118095.812  & -119275.437  & -119274.983  \\
    DeePKS-Cu & -587.964  & -118052.757  & -119233.078  & -119232.743  \\
    \multicolumn{5}{c}{Energies calculated at method-relaxed geometries}\rule{0pt}{12pt} \\
    HSE   & -587.974  & -118052.950  & -119233.227  & -119232.781  \\
    DeePKS-Rh & -587.976  & -118052.887  & -119233.193  & -119232.732  \\
    DeePKS-Cu+Rh & -587.975  & -118052.926  & -119233.241  & -119232.762  \bigstrut[b]\\
    \hline
    \hline
    \end{tabular}%
  \label{tab:addlabel}%
\end{table}%
